# Decoding *Dmrt1*: Insights into vertebrate sex determination and gonadal sex differentiation

Barbora Augstenová[1], Wen-Juan Ma[1*]

1. Research group of Ecology, Evolution and Genetics, Biology Department, Vrije Universiteit Brussel, Brussels, Belgium

*Corresponding: wen-juan.ma@vub.be

## Abstract

The *Dmrt* gene family is characterized by a conserved DM domain. It includes nine genes in vertebrates and is crucial to sex determination and sexual differentiation. *Dmrt1* is pivotal in testis formation and function by interacting with genes crucial for Sertoli cell differentiation, such as *Sox9*. *Dmrt1*, or *Sox9*, forms a conserved antagonistic interaction with *Foxl2* (crucial for ovarian formation) across mammals. Across 128 vertebrate species, *Dmrt1* exhibits sexually dimorphic expression, prior to and during gonadal sex differentiation and in adult testes, implicating its role in master regulation of sex determination and gonadal sex differentiation. *Dmrt1* emerges as a master/upstream sex-determining gene in one fish, frog, chicken and turtle, with candidacy in 12 other vertebrate species. Recent studies suggest epigenetic regulation of *Dmrt1* in its promoter methylation, and transposable element insertion introducing epigenetic modification to *cis*-regulatory elements of *Dmrt1*, alongside non-coding RNA involvement, in a wide spectrum of sex-determining mechanisms ranging from genetic factors, to interactions between genetic factors with the environment, to solely environmental factors. Additionally, alternative splicing of *Dmrt1* was found in all major vertebrate groups except amphibians. *Dmrt1* has evolved many lineage-specific isoforms (ranging from 2 to 10), and various isoforms showed sex, tissue or development-specific expression, which is in contrast to the highly conserved sex-specific splicing of its homolog *Dsx* across insects. Future research should focus on understanding the molecular basis of environmental sex determination from a broader taxon, and the molecular basis of epigenetic regulation. It is also essential to understand why and how multiple alternative splicing variants of *Dmrt1* evolve in vertebrates, the specific roles each isoform plays in sex determination and gonadal sex differentiation, as well as the significant differences in the molecular mechanisms and functions of alternative splicing between *Dmrt1* in vertebrates and *Dsx* in insects. Understanding the differences could provide deeper insights into the evolution of sex-determining mechanisms between vertebrates and insects.

## 1. The structure and evolution of *Dmrt* genes

The name of the *Dmrt* genes derives from the fruit fly *Drosophila melanogaster doublesex* (***D****sx*) gene and the nematode worm *Caenorhabditis elegans male abnormal 3 (**M**ab-3)* gene **R**elated **T**ranscriptional factors (Erdman & Burtis, 1993; Raymond *et al.*, 1998). These genes (*Dmrt* genes, *Dsx*, *Mab-3*) are characterized by the presence of the DM domain, a DNA-binding motif, which is an unusual cysteine-rich zinc DNA binding motif composed of two intertwined zinc fingers (Bellefroid *et al.*, 2013). These genes with DM domain are quite conserved in their role of regulating sexual development in arthropods, nematodes and chordates, and of directing gonadal development in all animals and controlling overall sexually dimorphic traits in a wide range of taxa (Matson & Zarkower, 2012; Zarkower, 2013). They are thought to have an ancient role in sex determination ancestral to all metazoan species (Matson & Zarkower, 2012). At the structural level, the DM domain contains a novel zinc module consisting of intertwined $Zn^{2+}$-binding sites, and a disordered tail which functions as a nascent recognization of α-helix. In *Drosopihla* flies, mutations in either $Zn^{2+}$-binding site or tail can lead to an intersex phenotype, and affect sexual dimorphism and courtship behavior (Zhu *et al.*, 2000). At the functional level, genes with DM domains are at the interface between sex determination and sexual differentiation. Genes with the DM domain integrate sexual differentiation with developmental patterning and regulate the downstream target genes in varied tissues and cell types (Matson & Zarkower, 2012). These genes often have functional pleiotropy, which constrained their evolution and were often prone to duplication and functional divergence. However, it remains largely unknown why and how the genes with DM domain can regulate sex at largely diverged groups across metazoans (Matson & Zarkower, 2012).

Different taxa have evolved different numbers of *Dmrt* genes (Bellefroid *et al.*, 2013; Wang *et al.*, 2023). In vertebrates, there are overall recognized nine *Dmrt* genes *Dmrt1 – Dmrt8* and *Dmrt2b* (Figure 1). The gene *Dmrt2b* is restricted to fishes. Earlier studies suggested the *Dmrt6* was missing in fishes, however, recent studies reported *Dmrt6* in certain fishes including coelacanth, tilapia, channel catfish and southern catfish (Forconi *et al.*, 2013; Zhang *et al.*, 2014; Wan *et al.*, 2020; Wang *et al.*, 2023). The genes *Dmrt7* and *Dmrt8* are only present in mammals (Figure 1) (Bellefroid *et al.*, 2013; Dong *et al.*, 2020). *Dmrt2, Dmrt3, Dmrt4, Dmrt5*, and *Dsx*-like genes were suggested to be widely distributed in vertebrates (Bellefroid *et al.*, 2013, Dong *et al.*, 2020). DMRT proteins have been classified into different subfamilies, based on additional conserved protein domains and on the exon-intron structures (Figure 1). The nomenclature of the *Dmrt* genes is not always clear, e.g. in mouse the genes *Dmrt3, Dmrt4, Dmrt5* are also known as *Dmrta3, Dmrta1, Dmrta2* respectively (Bellefroid *et al.*, 2013). *Dmrt3*, *Dmrt4* and *Dmrt5* belong to one subfamily characterized by the presence of a conserved DMA domain at their C-terminus, and each gene has two coding exons, with the DM domain encoded by the first coding exon and the DMA domain encoded by the second one (Bellefroid *et al.*, 2013, Sheng *et al.*, 2014). The structure of *Dmrt5* can be characterized additionally by a conserved DMB domain near the C-terminus (Figure 1) (Guo *et al.*, 2004; Sheng *et al.*, 2014). The DM domain of *Dmrt8* is not so conserved across mammals in comparison to other *Dmrt* genes and contains many mutations, and in some cases the functional DM domain was lost altogether (Figure 1) (Bellefroid *et al.*, 2013).

Regarding the evolutionary origin of *Dmrt* genes, no conclusive phylogenetic relationship among all DMRT proteins can be drawn at the level of metazoans, because there is little sequence similarity outside the DM and DMA domains (Bellefroid *et al.*, 2013), and the length of DM domain (~70 amino acids) was too short to provide phylogenetic certainty. However, a reduced dataset revealed eight monophyletic groups containing *Dmrt* genes, and most clades have phylogenetically limited distributions (Wexler *et al.*, 2014). For instance, *Dmrt1* orthologs were mainly found in vertebrates such as mammals, birds, amphibians and teleost fishes (Wexler *et al.*, 2014), suggesting *Dmrt1* might originate in the common vertebrate ancestor (Mawaribuchi *et al.*, 2019). Orthologs of *Dmrt2, Dmrt3*, *Dmrt4* and *Dmrt5* were already present in the ancestor of deuterostomes, containing both DM and DMA domains (Figure 1). Comparison between DM and DMA domain phylogenies suggests that vertebrate *Dmrt2* has lost its DMA domain since related sequences in cephalochordates and hemichordates contain this domain (Bellefroid *et al.*, 2013). The *Dmrt2* ortholog group, containing sequences from different taxa, appears to have diverse roles in various metazoan taxa such as somitogenesis and regulation of left-right axial patterning. *Dmrt8* shows strong similarity over their entire sequence to *Dmrt7*. Both have been found in mammals but not in other vertebrate species. In *Dmrt8* from human, chimp, orangutan and possibly other mammals, the open reading frame is disrupted by a stop codon 5' of the DM domain, hence they could not encode a functional DM domain (Veith *et al.*, 2006). Nevertheless the downstream of the DM domain can code for DM domain-less DMRT8 proteins, which could evolve as dominant-negative regulators to form heterodimers with other DMRT proteins preventing DNA binding (Veith *et al.*, 2006). In human, *Dmrt8* is expressed in various somatic tissues and gonads in both sexes (Ottolenghi *et al.*, 2002). In contrast, three isoforms were detected in each of two *Dmrt8* genes and all isoforms were expressed in embryonic testis and adult testis in the mouse (Veith *et al.*, 2006).

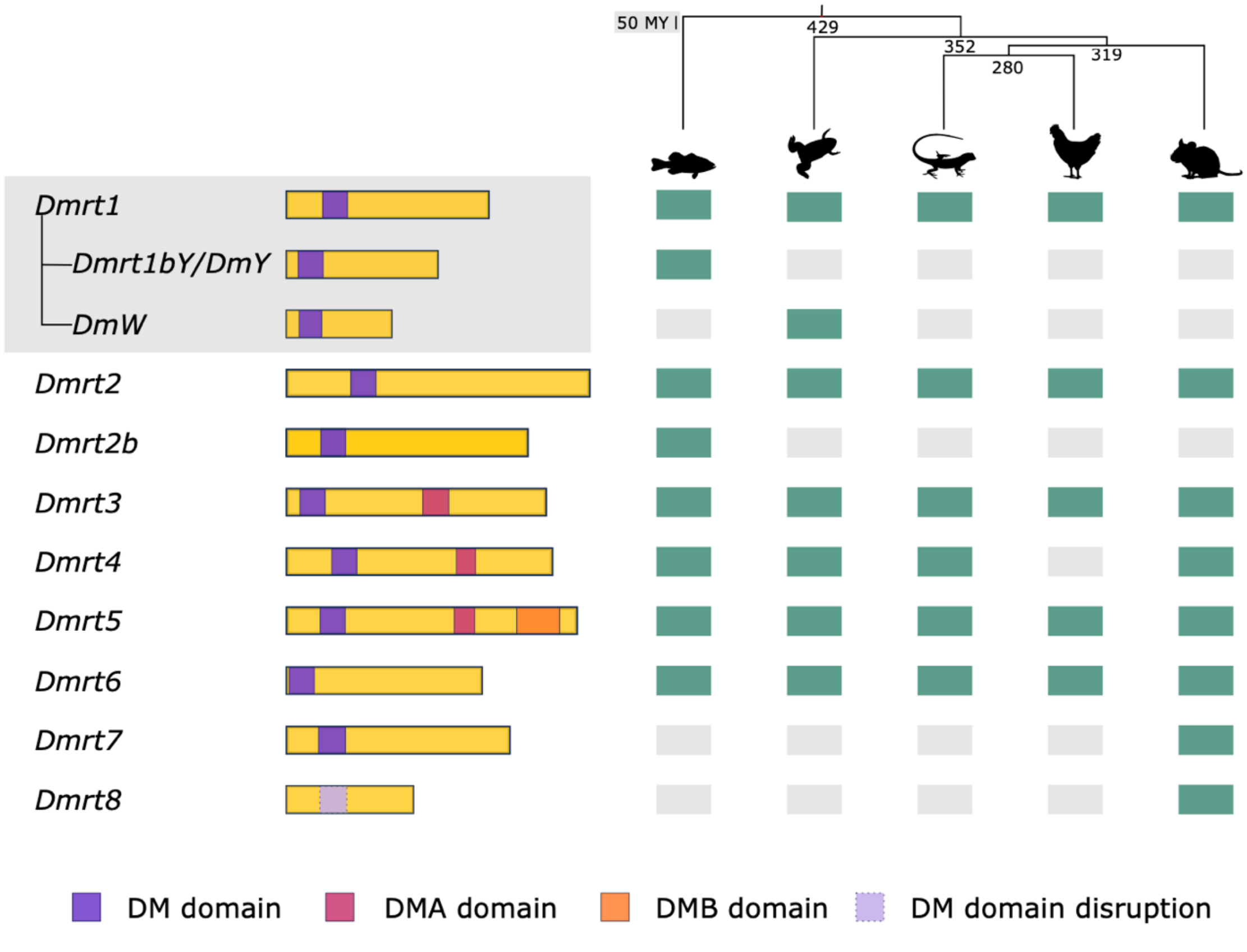

**Figure 1**. Domain architecture of *Dmrt* gene family and their main distribution across vertebrate groups. All *Dmrt* genes share the highly conserved protein motif DM domain. *Dmrt3*, *Dmrt4* and *Dmrt5* also share a highly conserved DMA domain at their C-terminus. *Dmrt5* has an additional conserved DMB domain. The conserved DM domain on *Dmrt8* was disrupted and thus produced non-functional DM domain (or DM domain less) protein. Two important key paralogs of *Dmrt1*, *Dmrt1bY*/*DmY* and *DmW*, are also included. These paralogs were identified as master sex-determining genes in the madaka fish *Oryzias latipes* (Matsuda *et al.*, 2002; Nanda *et al.*, 2002) and the tropical clawed frog *Xenopus laevis* respectively (for details see section 5 "*Dmrt1* and its paralogs' role as master sex-determining gene") (Yoshimoto *et al.*, 2008). The phylogenetic tree was obtained from the open source http://www.timetree.org.

## *2. Dmrt1* and its regulatory network

*Dmrt1* is indispensable for the maintenance of testis development and function, and the suppression of female-determining pathways in males across various vertebrate groups (Ferguson-Smith, 2007; Hong *et al.*, 2007). Its crucial role in sex determination has been documented in chicken, frogs, fishes and reptiles, where *Dmrt1* and its paralogs have been identified as the master, or candidate master, sex-determining gene (Table S1) (Matsuda *et al.*, 2002; Ferguson-Smith, 2007; Yoshimoto *et al.*, 2008; Smith *et al.*, 2009; Rhen & Schroeder, 2010; Herpin & Schartl, 2011; Ma *et al.*, 2016; Cui *et al.*, 2017; Ge *et al.*, 2017; Mustapha *et al.*, 2018). The highly conserved role of *Dmrt1* across vertebrates highlights its fundamental importance in the genetic regulatory networks that govern sex determination and gonadal sex differentiation.

The genetic regulatory network involving *Dmrt1* is complex yet essential for the proper development of testis and sexual differentiation in vertebrates. Most functional knowledge is from work in sexual development in mammals, especially in mice (Lindeman *et al.*, 2015, 2021; Rahmoun *et al.*, 2017). In mammals, gonadal sex determination is a binary switch where the biopotential somatic progenitor cells are triggered to differentiate into Sertoli cells in males or granulosa cells in females (Lindeman *et al.*, 2021). In mice, *Dmrt1* is necessary for testis maintenance, and is sufficient to induce female-to-male cell fate reprogramming *in vivo*. *Sox9*, another critical transcription factor for the determination of Sertoli cells that orchestrate testis development, works synergistically with *Dmrt1* to promote the differentiation and maintenance of Sertoli cells, which are essential for testis formation in vertebrates (Raymond *et al.*, 2000; Lindeman *et al.*, 2021). *Dmrt1* expression in female ovary cells at around the sex-determining period could suppress the *Foxl2* (forkhead transcription factor 2) expression which was on the top cascade of the female-determining pathway. This then activated expression of *Sox9, Sox8*, and reprogramed juvenile and adult (female) granulosa cells into (male) Sertoli-like cells which led to typical adult male Sertoli morphology (Lindeman *et al.*, 2015). *Dmrt1* was required continuously in pubertal and adult testis to suppress *Foxl2* expression and other regulators in female-determining pathway. In addition, *Dmrt1* also acted with *Sox8*/*Sox9* in female-to-male reprogramming. In Sertoli cells *Dmrt1* and *Sox9* bound many accessible chromatin sites, and *Dmrt1* could render postnatal granulosa cell chromatin accessible for *Sox9* binding (Lindeman *et al.*, 2021). Additionally, studies in mice on genome-wide DNA binding and transcriptional regulation, as well as with in vitro random oligonucleotide selection method, suggest *Dmrt1* could bind to its own promoter, indicating a possible autoregulatory feedback mechanisms (Murphy *et al.*, 2007, 2010).

The activity of *Dmrt1* could be modulated by various signaling pathways. For instance, the Notch signaling pathway plays a crucial role in the development and function of various tissues including the gonads. In mice, active Notch signaling in key cellular events took place at adult spermatogenesis, and blockade of Notch signaling disrupted the expression patterns of Notch components in the testis, induced male germ cell fate aberrations and increased germ cell apoptosis (Murta *et al.*, 2014). Furthermore, the pathway components interact with *Dmrt1* to fine-tune the sexual fate balance between male and female-determining pathways during embryonic gonadal development. In the *Dmrt1* mutant fetal testes, Notch signaling was highly disruptive, indicating certain interactions between *Dmrt1* and the Notch signaling components in the differentiation and proliferation of Sertoli cells which are essential for testis formation and spermatogenesis (Krentz *et al.*, 2013). Beyond Notch signaling, *Dmrt1* is also involved in crosstalk with the *Wnt* signaling pathway that regulates sexual differentiation. The *Wnt* signaling pathway is crucial for the formation and function of Sertoli cells and Leydig cells in mouse, which are secreted by steroid and located in the interstitial region between testis cords (Chassot *et al.*, 2008; Tang *et al.*, 2008). *Wnt4*-knockout mice developed ovotestes with the presence of sex cords and functional steroidogenic Leydig cells (Kim *et al.*, 2006; Chassot *et al.*, 2008). Double knockout of two ovarian somatic factors like *Wnt4* and *Foxl2*, resulted in limited upregulation of *Dmrt1* and strong upregulation of *Sox9*, leading to the formation of testis tubules and spermatogonia in XX (female) mice. This suggests the *Wnt* pathway interacts with *Dmrt1* and *Sox9* to regulate the balance between female and male gonadal differentiation (Kim *et al.*, 2006; Ottolenghi *et al.*, 2007; Chassot *et al.*, 2008).

## 3. *Dmrt1* expression during development and adulthood in vertebrates

*Dmrt1* is crucially involved in gonadal sex differentiation across a wide range of taxa in vertebrates (Mawaribuchi *et al.*, 2012; Ayers *et al.*, 2013; Huang *et al.*, 2017). The role of *Dmrt1* in testes development and functionality in vertebrates was first described in humans, where the deletion of autosomal chromosome 9 region containing *Dmrt1* (placed in the distal 9p region - 9p24.3) led to complete or partial gonadal dysgenesis, and hemizygosity of this region led to testicular defects and further to XY sex reversals (Veitia *et al.*, 1997; Raymond *et al.*, 1999a; Ledig *et al.*, 2010).

*Dmrt1* has diverse expression dynamics in the somatic and gonadal tissues during development in vertebrates. In Table S1 with 128 vertebrate species (76 in fishes, 17 in amphibians, 16 in mammals and 19 in reptiles including avians), we have summarized the *Dmrt1* temporal and spatial expression pattern in gonads prior to and during gonadal sex differentiation, at sub-adult and adult stages, as well as various adult somatic tissues. Several general patterns can be drawn from Table S1. First, in a handful of species, *Dmrt1* showed either similar or higher expression levels in undifferentiated XY gonads or in the whole XY embryos. Second, for the vast majority of species, *Dmrt1* showed a sexually dimorphic expression pattern during the gonadal sex differentiation period with increased expression in males and decreased in females, highlighting its conserved role involving testis formation during development. However, in certain frog species, such as *Physalaemus pustulosus*, *Pelophylax nigromaculatus*, *Rhinella marina* (formerly *Bufo marinus*), there was no detected difference in *Dmrt1* expression during gonadal sex differentiation period (Table S1) (Abramyan *et al.* 2009; Duarte-Guterman *et al.*, 2012; Xu *et al.*, 2015). This suggests *Dmrt1* may not be the main master sex-determining genes in the three studied frog species. Third, in sub-adult or adult gonads, *Dmrt1* is exclusively expressed in testes in the 27 studied species, has significantly higher expression in testes and weaker expression in ovaries in about 52 species, and has inconclusive expression pattern in 11 species due to expression data in testes but no data examined in ovaries (Table S1). *Dmrt1* expression is also detected in ovotestes with either weak expression in protandrous hermaphrodite fish *Acanthopagrus latus* (individuals first develop mature testes which later transition into ovaries), or stronger expression than either testes or ovaries in protogynous hermaphrodite fish *Epinephelus coioides* (individuals first develop mature ovaries which later transition into testis) (Table S1). This is consistent with *Dmrt1*'s role in formation and maintenance of testicular development. Lastly, *Dmrt1* has no expression or very weak expression in adult somatic tissues such as brain, liver, kidney, mussel, heart, spleen, eye, among others (Table S1).

*Dmrt1* expression changes are associated with sex change in certain fish species with sequential hermaphroditism. The *Dmrt1* expression in protandrous clownfish *Amphiprion bicinctus* (Pomacentridae) decreased in gonads during and after the transformation from male to female (Casas *et al.*, 2016). In the closely related clownfish species *Amphiprion clarkii*, *Dmrt1* was expressed in nonbreeder (juvenile males, that do not participate in mating and reproduction until either the dominant female or male dies to take over their role) and male gonads (Wang *et al.*, 2022a). However, *Dmrt1* expression level was retained and, surprisingly, at even higher levels in ovaries after the sex transformation, suggesting a potential role in promoting ovarian development and required further functional studies in this species (Casas *et al.*, 2016; Wang *et al.*, 2022a). In the fish species with protogynous hermaphroditism such as *Monopterus albus* (Synbranchidae), *Epinephelus merra* (Serranidae) or *Halichoeres poecilopterus* (Labridae), the expression of *Dmrt1* increased in gonads at the time of the transformation from females to males, suggesting a role of *Dmrt1* in testis formation and development (Huang *et al.*, 2005; Alam *et al.*, 2008; Miyake *et al.*, 2012).

As summarized in Table S1, studies have shown that *Dmrt1* at the adult stage is predominantly expressed in the testis. Interestingly, *Dmrt1* expression levels vary with breeding season conditions. Various studies across vertebrates show that *Dmrt1* expression differs between the breeding and non-breeding season with higher expression in the testis during the breeding season (Table S1), suggesting *Dmrt1* expression is associated with the active status of spermatogenesis. The higher expression of *Dmrt1* in testis of males in breeding season has been found in the Iberian mole *Talpa occidentalis* (Talpidae) (Dadhich *et al.*, 2010), in various rodent species (Massoud *et al.*, 2021), in reptiles such as the Italian wall lizard *Podarcis sicula* (Lacertidae) (Capriglione *et al.*, 2010), as well as the rainbow trout fish *Oncorhynchus mykiss* (Salmonidae) (Marchand *et al.*, 2000).

## 4. *Dmrt1's* role in gonadal sex differentiation in vertebrates

### 4.1 Function of *Dmrt1* in germ cell development

The transcriptional regulation function of *Dmrt1* is best studied in mammals. *Dmrt1* was expressed in the genital ridge and the gonadal primordium in both sexes, and studies of gene/allelic knock-in and knockout demonstrated that at least one functional copy of *Dmrt1* was required for the proliferation and radial migration of germ cells in the mouse. Germ cells died by about postnatal day 10 in *Dmrt1* mutant testes (Raymond *et al.*, 2000; Fahrioglu *et al.*, 2006; Kim *et al.*, 2007). Interestingly, functional studies showed that *Dmrt1* determined whether male germ cells (spermatogonia) had undergone mitosis and spermatogonia differentiation or entered meiosis, and was also required for normal meiotic prophase in the embryonic ovaries and formation of primordial follicles. For instance, loss of *Dmrt1* in spermatogonia disrupted gene expression in Sertoli cells, causing spermatogonia to exit the spermatogonial program and entered meiosis (Matson *et al.*, 2010, 2011). In addition, Sertoli cell-specific deletion of *Dmrt1* disrupted meiosis in prophase I, due to a lack of supporting cell function in the mutant Sertoli cells (Zarkower, 2013). Furthermore, loss of *Dmrt1* in the embryonic ovary resulted in the reduction of the number of primordial follicles in the juvenile ovary, although the females were still fertile (Krentz *et al.*, 2011). In rabbits, a recent study shows *Dmrt1* was involved in testis determination and was also essential for female fertility. The XX and XY embryos with the knockout of *Dmrt1* did not undergo meiosis and folliculogenesis, and XX adult females without *Dmrt1* were sterile (Dujardin *et al.*, 2023). Overall, *Dmrt1* is required for germ cell differentiation in males, regulates the initiation of meiosis in both sexes and is crucial for supporting cells (i.e. Sertoli cells) to initiate and maintain testis development in mouse, and for testes differentiation in therian mammals (Raymond *et al.*, 2000; Kim *et al.*, 2007; Zarkower, 2013). *Dmrt1* seems to be also indispensable in primordial follicles development and for female fertility at least in rabbits (Dujardin *et al.*, 2023).

**4.2 *Dmrt1*'s role in testis development**

Functional studies of *Dmrt1* in non-mammals have been constrained to only a handful of species. For instance, *Dmrt1* or its paralogues were identified as candidate master sex-determining genes in 7 fishes with knockdown or knockout functional assay, such as medaka (*Oryzias latipes*), *Astyanax scabripinnis*, Chinese tongue sole (*Cynoglossus semilaevis*), Siamese fighting fish (*Betta splendens*), spotted scat (*Scatophagus argus*), African scat (*Scatophagus tetracanthus*), spotbanded scat (*Selenotoca multifasciata*) (Table 1). For the rest of fishes, most species focused on expression profile during juvenile gonadal sex differentiation and at adult stages. In many fishes, such as Nile tilapia (*Oreochromis niloticus*), black porgy (*Acanthopagrus schlegeli*), spotted scat (*Scatophagus argus*), Chinese tongue sole (*Cynoglossus semilaevis*), spotted knifejaw (*Oplegnathus punctatus*), Japanese eel (*Anguilla japonica*), *Dmrt1* showed sexually dimorphic expression with higher expression in males during juvenile gonadal sex differentiation and was predominantly expressed in adult testis (He *et al.*, 2003; Li *et al.*, 2013; Chen *et al.*, 2014; Cui *et al.*, 2017; Mustapha *et al.*, 2018; Jeng *et al.*, 2019; Qi *et al.*, 2024; Zhao *et al.*, 2024). In the zebrafish (*Danio rerio*), knockdown of *Dmrt1* at the early embryonic stage resulted in fertile females, or in rare cases resulted in males with testicular dysgenesis, suggesting a role for male germ cell maintenance and proliferation, and no requirement for ovary development (Lin *et al.*, 2017; Webster *et al.*, 2017). In the Siamese fighting fish (*Betta splendens*) with XX/XY sex chromosome system, *Dmrt1* showed male-biased expression in undifferentiated gonads and adult testis, and the knockout of *Dmrt1* resulted in ovarian development in XY individuals (Wang *et al.*, 2022b).

In many studied frog species such as *Xenopus laevis, Bombina bombina, Bufo viridis, Hyla arborea, Glandirana rugosa, Rana arvalis and Rana temporaria*, *Dmrt1* showed strong sexually dimorphic, male-biased expression in gonads either throughout development (from early embryos to late froglet stages where phenotypic gonadal sex differences can be confidently distinguished under the microscope) and/or at adult stage (Table S1) (Shibata *et al.*, 2002; Piprek *et al.*, 2013; Ma *et al.*, 2018a; b). In the European common frog *Rana temporaria*, *Dmrt1* is located on the sex chromosome. Population genetics and pedigree analysis showed that $F_{st}$ (the proportion of the total genetic variance in a subpopulation relative to the total genetic variance. High value implies a considerable genetic differentiation among groups) between sexes was only significantly elevated at the *Dmrt1* region along the entire linkage group 2 (sex chromosome), and *Dmrt1* polymorphism (*Dmrt1* Y-chromosome haplotypes) covaried with sex determination in various populations across Europe (Ma *et al.*, 2016; Rodrigues *et al.,* 2017). Furthermore, *Dmrt1* Y-chromosome haplotype was also associated with varying levels of masculinization, with evidence of association with male testis formation and development throughout development (Rodrigues *et al.*, 2017; Ma *et al.*, 2018a; b). All evidence together points out *Dmrt1* is a strong candidate master sex-determining gene in this species (Table 1). In the salamander

*Hynobius retardatus,* high temperature led to downregulation of *Dmrt1* expression, resulting in the development of sex reversals (Sakata *et al.*, 2006).

Most reptiles have genetic sex determination (GSD). All avians possess GSD with ZZ/ZW system, and Squamata mostly have GSD with either XY/XX or ZW/ZZ systems (Sarre *et al.*, 2011). However, sex, in certain lineages such as turtles and crocodiles, is predominantly determined by temperature (TSD) in the thermo-sensitive period during embryogenesis, which is one primary trigger of environmental sex-determining mechanism (ESD) (Table S1) (Valenzuela & Lance, 2004; Bista & Valenzuela, 2020). Interestingly, in the Australian bearded dragon, GSD (ZW/ZZ system) was sensitive to temperature and was overturned by temperature within one generation. The study showed instant creation of a lineage entirely composed of ZZ bearded dragons when altering the incubation temperature during thermo-sensitive period, and W chromosome was eliminated in these animals (Holleley *et al.*, 2015). In most studied TSD species, e.g. *Alligator mississippiensis, Lepidochelys olivacea, Chelydra serpentina, Trachemys scripta*, *Dmrt1* was expressed and increased during temperature-sensitive embryogenesis period at male-producing temperature (Table S1) (Rhen *et al.*, 2007; Urushitani *et al.*, 2011; Bieser & Wibbels, 2014; Ge *et al.*, 2017; Diáz-Hernández *et al.*, 2020), which was consistent with the role of *Dmrt1* involving in testis formation and function. Furthermore, *Dmrt1* was suggested to play an important role in proliferation and maintenance of spermatogonia and spermatocytes in the Italian wall lizard *Podarcis sicula* (Capriglione *et al.*, 2010).

## 5. *Dmrt1* and its paralogs' role as master sex-determining gene

The DM domain has a conserved role in sexual development across both vertebrates and invertebrates (Hong *et al.*, 2007; Van De Zande & Verhulst, 2014; Zarkower & Murphy, 2022), which makes *Dmrt1* a good candidate for the master sex-determining gene, as well as a crucial component of the sex-determining pathway in many vertebrates. So far *Dmrt1*, or its paralogs *Dmrt1bY* or *DmY* and *DmW* (Figure 2), are demonstrated to be the master sex-determining genes in chicken, fish and frog respectively (Matsuda *et al.*, 2002; Nanda *et al.*, 2002; Yoshimoto *et al.*, 2008, 2010; Smith *et al.*, 2009), and are candidate master sex-determining genes in 12 other vertebrates (Table 1). In contrast to the high diversity of master sex-determining genes, the downstream genetic cascades of (gonadal) sexual differentiation are relatively conserved within and to a certain extent across vertebrate groups (Matson & Zarkower, 2012).

### 5.1 *Dmrt1bY/DmY* as master sex-determining gene in medaka fish

Duplication of *Dmrt1*, *Dmrt1bY* or *DmY*, was identified as master sex-determining gene in the medaka fish (*Oryzias latipes*) (Figure 2) (Matsuda *et al.*, 2002; Nanda *et al.*, 2002). The same gene was later found to be involved in sex determination in one closely related species *O. curvinotus*, but not in any other closely related fish species (Matsuda *et al.*, 2002; Zhang, 2004). Both species possess XX/XY sex chromosomes and the master sex-determining gene *Dmrt1bY/DmY* originated from *Dmrt1* by duplication in the common ancestor ca 4-10 million years ago (Mya) (Nanda *et al.*, 2002; Kondo *et al.*, 2004).

The *Dmrt1bY/DmY* gene plays a key role in testis development, and it alone is sufficient to lead to testes differentiation despite the exact regulatory network for the pathway is not well understood (Figure 2) (Matsuda *et al.*, 2002; Kobayashi *et al.*, 2004; Li *et al.*, 2022). *Dmrt1bY/DmY* also played an important role in regulating the primordial germ cells proliferation during embryonic gonad development and was expressed at gonadal undifferentiated stage (Figure 2). The *Dmrt1* itself was expressed later after the testis differentiation was completed (Kobayashi *et al.*, 2004; Herpin *et al.*, 2007). In mature testes, both *Dmrt1* and *Dmrt1bY/DmY* were expressed in Sertoli cells. *Dmrt1bY/DmY* expression was lower in mature testes, yet *Dmrt1* expression was relatively higher and was also detected in the testis of XX sex-reversed males, indicating *Dmrt1* is crucial in spermatogenesis in medaka fish. Together it suggests separate roles for each gene, *Dmrt1bY/DmY* for testis formation and differentiation and *Dmrt1* for spermatogenesis (Kobayashi *et al.*, 2004). Interestingly, examination of one chimeric fish where germ cells consisted of XX cells but the somatic cells of the testis contained both female (XX) and male (XY) cells, indicating that XY cells with *Dmrt1bY/DmY* could be sufficient for testis development (Shinomiya *et al.*, 2002). Neither *Dmrt1* nor *Dmrt1bY/DmY* was expressed in ovaries (Kobayashi *et al.*, 2004). Within teleost fishes, for the conserved downstream cascade of gonadal sex differentiation, the testis development usually consists of the interactions among *Dmrt1*, *Sox9, Amh* (*anti-müllerian hormone*) and *Gsdf* (*gonadal soma-derived factor*), while ovary development contains *Foxl2* and *Cyp19a1* (*cytochrome P450 family 19 subfamily a member 1*) (Figure 2) (Li *et al.*, 2022).

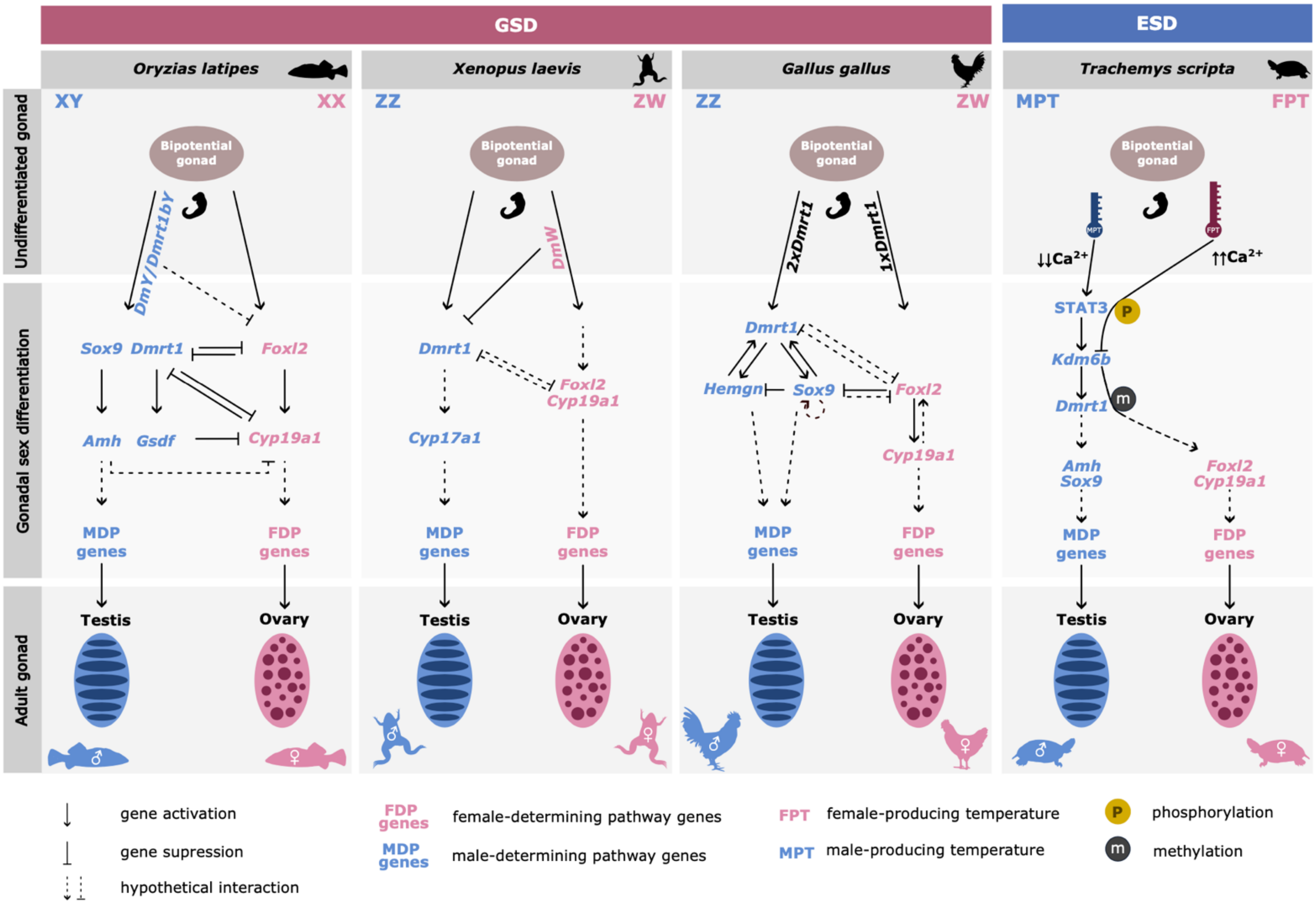

**Figure 2**. Diagram of the key sex-determining pathway of *Dmrt1* and its paralog *Dmrt1bY* or *DmY, DmW*, being characterized as the master sex-determining gene in medaka fish, African clawed frog, chicken and crucial upstream gene in the red-eared slider turtle. In vertebrates, all gonads start developing as undifferentiated gonads and then gonadal sex differentiation progresses. The master sex-determining gene *Dmrt1* and its paralogs initiate their expression either prior to or at the onset of gonadal sex differentiation during early embryogenesis. The main gene network directly regulated by *Dmrt1* and its paralogs are illustrated in each case which ultimately leads to differentiated gonads at the adult stage. Gene regulator network illustration is adapted from previous studies for fish (Li *et al.*, 2022), frog (Ito, 2018), chicken (Sánchez & Chaouiya, 2018) and turtle (Weber *et al.*, 2020; Li *et al.*, 2022) respectively.

Spontaneous XX sex-reversed males are not uncommon in medaka fish. XX individuals can develop as fully functional males even without the presence of *Dmrt1bY/DmY* due to the presence of yet-to-be identified autosomal modifiers (Nanda *et al.*, 2003). Noticeably, at 34°C all XX individuals would develop as XX sex-reversed males, indicating temperature as an important factor in inducing XX sex reversals in medaka (Nanda *et al.*, 2003). Interestingly, no XY sex reversals (XY females) were found at low temperature treatment, suggesting *Dmrt1bY/DmY* expression was not affected by temperature. On the other hand, high temperature also led to the expression of *Dmrt1* during early embryonic development (Hattori *et al.*, 2007), indicating an important role of *Dmrt1* in the testis development in the XX sex-reversed individuals. Taken together, *Dmrt1bY/DmY* is sufficient, however not essential, to induce male pathway development. On the other hand, autosomal copy of *Dmrt1* seems crucial for the development and maintenance of functional testis (Masuyama *et al.*, 2012).

### 5.2 *DmW* as master sex-determining gene in the African clawed frog

A partial duplication of *Dmrt1*, *DmW*, located on the W chromosome, was dominant and female specific, and identified as the master sex-determining gene in the African clawed frog *Xenopus laevis* (Figure 2) (Yoshimoto *et al.*, 2008, 2010). Due to allopolyploidization in the ancestral species, *X. laevis* possesses two copies of *Dmrt1*, *Dmrt1.L* (*DMRT1α)* and *Dmrt1.S (DMRT1β*), located on the autosomal chromosome 2L and 2S, respectively (Yoshimoto *et al.*, 2008; Bewick *et al.*, 2011). *DmW* was the result of partial duplication of *Dmrt1.S (Dmrt1β),* maintaining the high similarity of 5'-coding region (exon1-4) but lacking the 3'-coding portion of transactivation domain-coding region (exon5-6) (Yoshimoto *et al.*, 2008, 2010; Mawaribuchi *et al.*, 2017a). The *Dmrt1* duplication occurred approximately 47 Mya after allotetraploidization, and *DmW* was found in several closely related *Xenopus* species (Bewick *et al.*, 2011; Mawaribuchi *et al.*, 2017b; Cauret *et al.*, 2020, 2023). Other than *X. laevis*, *DmW* was amplified in several females but none of the males in *X. gilli* using targeted next-generation sequencing, suggesting female-specific in this species, but was not consistently female-specific in most other *Xenopus* species (Cauret *et al.*, 2020, 2023). It is possible that after allotetraploidization *DmW* evolved within certain *Xenopus* lineage, but was independently lost, or segregated autosomally, or evolved other functions than sexual differentiation in most species except in *X. laevis* and *X. gilli* (Cauret *et al.*, 2020, 2023). The functional role of *DmW* in female gonad differentiation of *X. gilli* is yet to be investigated.

*DmW* is required for ovary development in females in *X. laevis*, and it functions as an antagonistic competitor of *Dmrt1* (Figure 2) (Yoshimoto *et al.*, 2010). Both *Dmrt1* and *DmW* bind to the same regulatory regions of target genes (yet to be discovered) that are crucial in testes development. During the embryonic gonad development, *DmW* mRNA was more abundant than *Dmrt1* mRNA in the primordial gonads of ZW tadpoles early in sex determination, which could suppress the male-determining pathway by antagonizing *Dmrt1* transcriptional activity to prevent the target gene interacting with *Dmrt1*, leading to *Dmrt1* transcriptional repression (Figure 2). As a result, *DmW* inhibits the gene cascades for testis formation, resulting in ovary formation. *DmW* is missing in ZZ embryos and *Dmrt1* could bind to the target gene to initiate the transcriptional activity, which directs the sex-determining pathway towards testes development (Yoshimoto *et al.*, 2010; Yoshimoto & Ito, 2011). Interestingly, the WW individuals evolve as phenotypically normal females, suggesting that the Z-specific region on the Z chromosome is not required for the development of females (Mawaribuchi *et al.*, 2017b), which further confirms the major role of *DmW* in sex determination of *X. laevis*. However, it is possible that there are factors functioning upstream to *DmW* that have yet to be discovered (Yoshimoto *et al.*, 2008; Cauret *et al.*, 2023). Other than the master sex-determining gene *DmW*, the downstream cascade of sexual differentiation is little known in frogs. A model for sex determination and gonadal sex differentiation suggests that *DmW* expression results in high expression of *Foxl2* and *Cyp19a1*, the latter produces estrogen hormone leading to ovary formation in the female-determining pathway. On the other hand, without *DmW* expression, *Dmrt1* expression seems to trigger the expression of *Cyp17a1*, a key enzyme in the production of male sex steroids, eventually resulting in testis development (reviewed in Ito, 2018).

### 5.3 *Dmrt1* as master sex-determining gene in chicken

*Dmrt1* is located on the Z chromosome of chicken (*Gallus gallus*), which led to the speculation of the role of *Dmrt1* in birds' sex determination for years (Smith *et al.*, 1999, 2007; Nanda *et al.*, 2000; Ellegren, 2001). The function of *Dmrt1* in gonadal differentiation was finally confirmed with *Dmrt1* knockdown in ZZ (normally male) chicken embryos, which led to feminization with ovarian development instead (Smith *et al.*, 2009). *Dmrt1* is thus required for proper testis formation and function in chicken (Smith *et al.*, 2009). Over-expression of *Dmrt1* in ZW females led to increased *Sox9* (possibly *Amh* too) expression and induced male sex-determining pathway (Figure 2). This further antagonized the female pathway by reducing the expression of the critical feminizing enzyme aromatase in embryonic chicken gonads (Lambeth *et al.*, 2014).

Because birds lack global Z chromosome-level dosage compensation, and chicken embryos need two functional copies of *Dmrt1* in order to develop into males with functional testes (Smith *et al.*, 2009; Lambeth *et al.*, 2014), it was proposed that *Dmrt1* was an essential sex-linked regulator of gonadal differentiation in the chicken and probably across avians, and it likely acts with a dosage mechanism (Lambeth *et al.*, 2014). Knockout of one *Dmrt1* allele in ZZ embryos, with CRISPR-Cas9-based monoallelic targeting approach, led to ovaries development in place of testes in chromosomally male (ZZ) chicken (Ioannidis *et al.*, 2021). In experiments where estrogen synthesis was blocked in ZW embryos, the resulting gonads developed as testes when *Dmrt1* was in but as ovaries when *Dmrt1* was knocked out. This demonstrated the avian sex-determining mechanism is indeed based on *Dmrt1* dosage, but the estrogen production was linked to *Dmrt1* expression and was a key factor in primary sex determination in chicken (Ioannidis *et al.*, 2021). Interestingly, in contrast to mammals where their secondary sexual phenotype strongly depends on the consistent sex hormone environment in gonads, the *Dmrt1*-knockout ZZ adult chicken (with ovaries) displayed identical secondary sexual characteristics to the wild-type ZZ adult males, supporting so-called cell-autonomous sex identity in birds (Ioannidis *et al.*, 2021). Additionally, *Dmrt1* knockout in ZZ male chicken embryos led to incomplete testis feminization, and the genome-wide gene expression analysis identified certain genes that could be involved in gonad feminization (Lee *et al.*, 2021). Since all birds possess homologous sex chromosomes and higher expression of *Dmrt1* prior to gonad differentiation was described not only in chicken but also in zebra finch and emu (Hirst *et al.*, 2017), it is thus possible that *Dmrt1* is required for gonadal sex differentiation in most birds (Table S1). Last but not least, regarding birds sex determination, the alternative scenario is that a female-specific dominant gene on the W chromosome functions as an upstream regulator to suppress *Dmrt1* gene that is yet to be discovered. It is relatively known for certain part of the downstream cascade of sex determination in chicken (reviewed in Sánchez & Chaouiya, 2018). For instance, in the male-determining pathway *Hemgn* upregulates *Dmrt1* which in turn activates *Sox9*, which provides feedback to *Dmrt1* to maintain its high expression, and possibly *Sox9* suppresses expression of *Hemgn*. *Dmrt1* further directly suppresses the expression of *Foxl2* and *Cyp19a1*. The regulatory network eventually results in testis development. On the other hand, in the female-determining pathway of no suppression from *Dmrt1*, *Foxl2* and *Cyp19a1* are expressed. Both could activate the Aromatase which is an enzyme for oestrogen synthesis in ovary formation and lead to ovary development ( reviewed in Cutting *et al.*, 2013; Sánchez & Chaouiya, 2018).

### 5.4 *Dmrt1* as an essential upstream regulator of temperature sex determination in turtle

The molecular mechanism of ESD has been a long-standing mystery until a recent discovery in the red-eared slider turtle *Trachemys scripta*. Prior to the study, the thermosensitive genetic triggers for gonadal sex differentiation remained largely unknown. Recent studies found the *Dmrt1* expression was temperature-dependent and showed a sexually dimorphic pattern preceding gonadal sex differentiation, and gene function analysis demonstrated *Dmrt1* was necessary and sufficient for initiation of testes formation and development (Kettlewell *et al.*, 2000; Ge *et al.*, 2017). The findings suggest that *Dmrt1* is an important upstream regulator of temperature-dependent sex determination (Ge *et al.*, 2017). *Dmrt1* expression was further regulated by the methylation in its promoter region (Ge *et al.*, 2018), and it is repressed in female-producing temperatures by phosphorylated signal transducer and activator of transcription 3 (*STAT3*), the phosphorylation was induced by calcium signaling pathways induced by high temperatures (Weber *et al.*, 2020). Like most non-model vertebrate systems, downstream cascade of sex determination is largely unknown. It has been hypothesized that *Dmrt1* regulates the expression of *Amh* and *Sox9* in the male-determining pathway, but suppresses the expression of *Foxl2* and *Cryp19a1* in the female-determining pathway (reviewed in Li *et al.,* 2022).

## 6. Role of epigenetic regulation of *Dmrt1* in sexual differentiation

Epigenetics is the study of changes in gene function that cannot be explained by changes in DNA sequences (Russo *et al.*, 1996). There are three main epigenetic regulation mechanisms: DNA methylation, modification of histones, and the involvement of non-coding RNAs (ncRNAs) (Piferrer, 2013). Epigenetic regulation allows organisms to integrate environmental and genetic factors to modify their gene activities for a particular phenotype, and play an essential role in various biological processes across the tree of life, such as germline imprinting, X-chromosome inactivation in mammals, mating-type silencing in yeast, temperature-dependent vernalization in plants (Piferrer, 2013, 2021; Gunes *et al.*, 2016; Tachibana, 2016).

Sex determination is typically categorized as genetic (GSD) or environmental mechanism (ESD), where genetic (gene or genes) or environmental (temperature, PH, social cues, salt etc.) factors function as the primary signal of the pathway to develop as a female or a male (Beukeboom & Perrin, 2014). However, sex determination in many species depends on the contribution of both genetic and environmental factors. It has been widely accepted that sex determination is a spectrum between pure GSD and ESD, where GSD and ESD only represent the opposite ends of the continuum (Figure 3, Beukeboom & Perrin, 2014; Straková *et al.*, 2020). Epigenetic regulation of sex determination strictly is not equal to ESD, because the epigenetic mechanisms such as DNA methylation, histone modification and ncRNAs can be involved in the whole spectrum of sex determination from GSD to ESD regardless of the primary triggering factors. For instance, epigenetic regulation can take place in GSD, or in systems where GSD interacts with ESD, or pure ESD species (Figure 3). The molecular basis of the sex-determining pathway of GSD has been extensively studied and in some cases is well understood (Figure 2), however, the underlying genetics of ESD are largely unknown. Furthermore, the role of epigenetic regulation on both GSD and ESD has not been well studied and more efforts thus should be carried on in such directions.

In this section, we delve into the emerging role of epigenetic regulation in sex determination, particularly focusing on the key sex-determining gene *Dmrt1*. While *Dmrt1*'s canonical function in orchestrating gonadal sex differentiation is well-established, the epigenetic regulation of *Dmrt1* and its implications for sex determination represent a novel frontier in the field. Here, we synthesize current knowledge and recent discoveries regarding the epigenetic regulation of *Dmrt1*, offering insights into its multifaceted role in sexual differentiation across diverse vertebrate taxa. More important question is to further identify the upstream epigenetic regulator responding to environmental factors, so far it has only been unraveled in the red-eared turtle species (see details in the next section). By exploring the epigenetic landscape of sex determination, we aim to shed light on the dynamic interplay between genetic and environmental factors underlying sex-determining pathways.

### 6.1 Epigenetic regulation of *Dmrt1* in species with ESD

To our best knowledge, the only study that demonstrates the molecular mechanism of ESD is in the red-eared slider turtle (*Trachemys scripta*) (Figure 2 and 3) (Ge *et al.*, 2018; Georges & Holleley, 2018; Weber *et al.,* 2020). This species has temperature-dependent sex determination (TSD), where embryos exposed to low temperatures during the thermo-sensitive developmental period develop as males, while at high temperatures they develop as females. Earlier study showed that the histone H3K27 (H3 lysine 27) demethylase KDM6B demonstrated sexually dimorphic expression, which was temperature-dependent during the early embryonic stage before the gonadal sex differentiation (Figure 2) (Ge *et al.*, 2018). Knockdown of *Kdm6b* at low temperature (i.e. 26℃, typically develop as males) led to male-to-female sex reversals in most surviving embryos. When the knockdown of *Kdm6b* and overexpression of *Dmrt1* were simultaneously performed, overexpression of *Dmrt1* could rescue the male pathway of *Kdm6b*-deficient gonads, suggesting KDM6B directly regulated the transcription of *Dmrt1* (by removing the trimethylation of H3K27 near the promoter of *Dmrt1*) (Ge *et al.*, 2018). Finally, a recent study identified a link between temperature and the activation of signal transducer and activator of transcription 3 (*Stat3*), which regulated the expression of *Kdm6b* (Weber *et al.*, 2020). They found that at warmer, female-producing temperature, an influx of $Ca^{2+}$ was initiated, which led *Stat3* to be phosphorylated that silenced *Kdm6b* transcription. The silencing of *Kdm6b* resulted in suppression of *Dmrt1* expression, which ultimately repressed testis development (Figure 2) (Weber *et al.*, 2020). Taken together, the study found the epigenetic mechanism is responsible for TSD by regulating *Dmrt1* expression at the early embryonic stages. The next step would be to investigate more ESD species to understand how broad this epigenetic mechanism is, before a general molecular mechanism for ESD can be drawn.

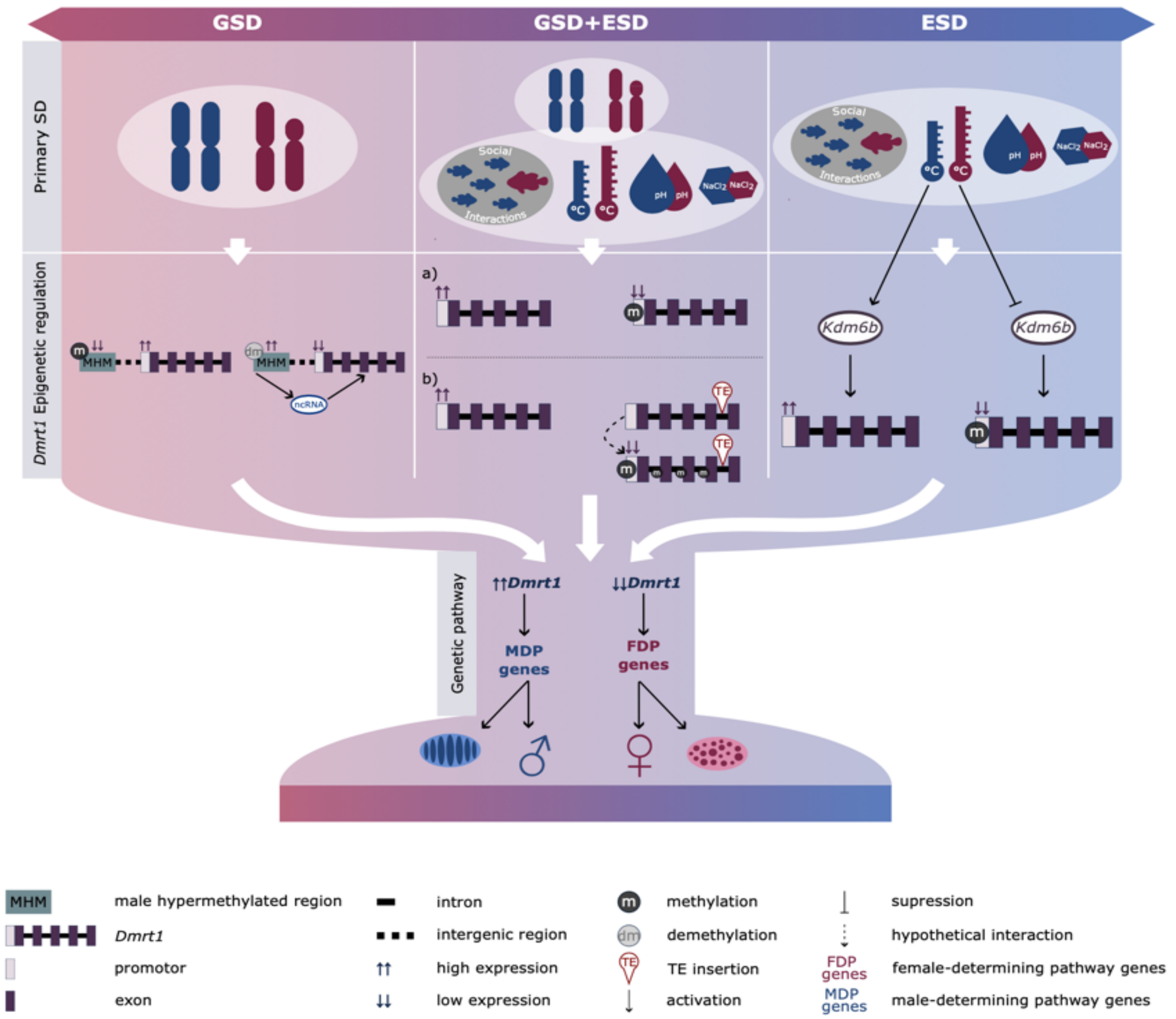

**Figure 3**. The model of epigenetic regulation of *Dmrt1* involved in sex determination and gonadal sex differentiation across the spectrum of sex determination, from GSD to ESD, is illustrated. The primary signals in sex determination range from GSD to ESD. Epigenetic regulation mechanism has been found in all spectrums of sex-determining systems, GSD, GSD interacts with ESD and ESD. In the GSD in the chicken, it was shown that the MHM region (male hypermethylated region) is located next to the *Dmrt1*. MHM is hypermethylated and transcriptionally inactive, which leads to upregulation of *Dmrt1*, resulting in testis development in males. In females, MHM is hypomethylated, transcribed to the ncRNA which downregulates *Dmrt1*, resulting in ovary development. In many fishes with GSD interaction with ESD, the current epigenetic regulation has found methylation of *Dmrt1* promotor region to suppress its expression, or involvement of transposable elements associated with methylation mechanism, where the exact molecular basis is yet to be discoverd.

### 6.2 Epigenetic regulation of *Dmrt1* in species where genetic interact with environmental factors

#### *6.2.1 Via methylation of Dmrt1 promoter region*

The epigenetic mechanism of sex determination has been also beautifully illustrated in a flatfish with sex determination involving interaction between genetic and environmental factors (Figure 3). In the Chinese half-smooth tongue sole (*Cynoglossus semilaevis*) with female heterogametic (ZZ/ZW) sex chromosome system, the disruption of spermatogenesis in *Dmrt1*-deficient ZZ males led to ovary-like gonad development, suggesting *Dmrt1* was an essential male sex-determining gene (Chen *et al.*, 2014; Zhou & Chen, 2016; Cui *et al.*, 2017; Wang *et al.*, 2019). Elevated temperature led to hypomethylation of the *Dmrt1* promoter in ZW embryos, which increased the *Dmrt1* expression at the time of sex determination, resulting in ZW pseudo-male development (Chen *et al.*, 2014). Furthermore, the methylation change was shown to be inherited across generations (Chen *et al.*, 2014). All these together demonstrate that temperature functions as stimuli to alter *Dmrt1* gene expression via epigenetic methylation mechanism to lead to the development of sex reversals. In the Nile tilapia (*Oreochromis niloticus*) with male heterogametic (XX/XY) sex chromosome system, functional studies show that once *Dmrt1* was mutated, the gonad could not be rescued to develop functional testis (Qi *et al.*, 2024). Several DNA methyltransferases (*Dnmts*) genes, responsible for DNA methylation, showed higher expression in the testis than in the ovary, and increased significantly in the (female to male) sex reversals than the control females. The *Dnmts* expressions were suppressed in methyltransferase-inhibitor treated gonads and *Dmrt1* expression was increased significantly, suggesting a role of epigenetic regulation in male testis development via modification of *Dmrt1* (Wang *et al.*, 2018). However, it is unclear what is the genetic mechanism for the methylation genes in controlling the *Dmrt1* expression in the Nile tilapia fish.

Other than epigenetic regulation of *Dmrt1* directly affects sex determination and leads to sex reversals in the aforementioned studies, it also shows that epigenetic modification acts on the *Dmrt1* expression to affect gonadal sex differentiation. For instance, in the Asian ray-finned fish (*Culter alburnus*), one study found that CpG methylation pattern of the *Dmrt1* promoter was highly sexually dimorphic, where it was hypermethylated in the ovaries but not in the testis. This leads to very high *Dmrt1* expression in testis, and almost non-detected in ovaries and several other somatic tissues (Jia *et al.*, 2019). In the Japanese flounder (*Paralichthys olivaceus*), 57-69% *Dmrt1* promoter CpGs was methylated in the ovary but not in the testis, leading to *Dmrt1* expression 70 times higher in the testis than in the ovary (Wen *et al.*, 2014). In the European sea bass (*Dicentrarchus labrax*), studies show sex-specific methylation and temperature response for *Dmrt1* and *Cyp19a1a* expressions in juvenile gonads, suggesting epigenetic regulation is possibly involved in gonadal differentiation in this species (Anastasiadi *et al.*, 2018). In the blotched snakehead (*Channa maculata*), the CpG methylation level of the *Dmrt1* promoter was correlated negatively with its gene expression in the testis and ovary, with *Dmrt1* being expressed 15 times higher in the testis than in the ovary during gonadal development. Steroid treatment affected *Dmrt1* expression and could induce male-to-female sex reversal. This study suggests *Dmrt1* is involved in testis differentiation and spermatogenesis in this species and epigenetic regulation can explain the gonadal differentiation throughout development (Ou *et al.*, 2022).

#### *6.2.2 Involving transposable elements (TEs)*

In the Siamese fighting fish (*Betta splendens*), the association mapping study shows that *Dmrt1* was located in the identified small sex-determining region and demonstrated a functional role where knockout of *Dmrt1* led to ovarian development in individuals with XY genotypes (Wang *et al.*, 2022b). Furthermore, this study found that a transposable element *Drbx1* inserted in the fourth intron of X-linked *Dmrt1* copy, which was associated with epigenetic modifications to the neighboring *cis*-regulatory elements in the *Dmrt1* promoter region, confirmed with methylation assays (Figure 3) (Wang *et al.*, 2022b). Similarly like in other fishes described above, this study again shows that methylation on *Dmrt1* protomer regions regulates the gene expression and leads to sex reversal, this time via TEs insertion with a yet unknown mechanism.

### 6.3 Epigenetic regulation of *Dmrt1* in species with GSD

*Dmrt1* was identified as the master sex-determining gene in chickens, and two copies of *Dmrt1* expression were required during early embryogenesis to initiate testis development (Smith *et al.*, 2009). A region neighboring *Dmrt1* was differentially methylated between males and females. In males, this region was hypermethylated, male hypermethylated region (MHM), and was transcriptionally inactive, whereas this region in females in hypomethylated and transcribed into a long ncRNA (Teranishi *et al.*, 2001). MHM is hypothesized to inhibit *Dmrt1* expression in females, as the MHM transcript accumulates at the site of transcription adjacent to *Dmrt1*. A functional study with injection of MHM into rooster testicles led to a strong downregulation of *Dmrt1* and a paling of the comb color, suggesting a change in sex hormones (Figure 3). Other studies show other sex-determining genes such as *Amh* and *Sox9* were unaffected at transcriptional levels, suggesting that MHM specifically regulates *Dmrt1* (Figure 3) (Yang *et al.*, 2010). Currently, the MHM has been primarily studied in chickens. Broader studies are needed to determine whether the proximity of the MHM to the *Dmrt1* region and the speculated regulation mechanism are chicken specific or a conserved epigenetic mechanism across avian lineages.

## 7. Role of alternative splicing of *Dmrt1* in gonadal sex differentiation

Alternative splicing refers to the mechanism where different exons or a fraction of an exon are joined in different combinations, to produce unique but related mRNAs transcripts, from a single pre-mRNA (Wang *et al.*, 2015a). Alternative splicing leads to an increase in the coding capacity of genes and therefore an increase of proteome complexity, constituting a ubiquitous regulatory mechanism of gene expression. Thus, alternative splicing was considered one key mechanism to contribute to phenotypic diversity among animal and plant taxa (Wright *et al.*, 2022). One of the most fundamental developmental processes, to which alternative splicing contributes, is sex determination (Wright *et al.*, 2022). The essential role of alternative splicing on sex determination has been extensively studied in a shared common pathway composed of the *Transformer (Tra)* gene and its downstream target *Doublesex* (*Dsx*) in insects (Suzuki *et al.*, 2001; Pomiankowski *et al.*, 2004; Cho *et al.*, 2007; Beukeboom & Zande, 2010). Female spliced *Tra* transcripts give rise to functional TRA proteins that directly regulate the splicing of female *Dsx* mRNAs, producing female DSX proteins and resulting in female development. When *Tra* is spliced into the male variant, no TRA proteins are produced. This results in the splicing of male isoform *Dsx* mRNAs, male DSX proteins and male development (Figure 4A) (Verhulst *et al.*, 2010; Gempe & Beye, 2011). Remarkably, sex-specific splicing of *Tra* and *Dsx*, key regulation genes in sex-determining and sexual differentiation in *Drosophila melanogaster*, is conserved across diverse insects, including the silk moth *Bombyx mori*, red flour beetle *Tribolium castaneum*, olive fruit flies *Bactrocera oleae*, the housefly *M. domestrica*, the *medfly Ceratis capitatia*, the honeybee *Apis mellifera* and the jewel wasp *Nasonia vitripennis*, suggesting an ancient origin and role in sex determination and sexual differentiation (Cho *et al.*, 2007; Giebel *et al.*, 2009; Verhulst *et al.*, 2010; Geuverink & Beukeboom, 2014; Van De Zande & Verhulst, 2014). *Dmrt1* is the homolog of *Dsx* in vertebrates, suggesting a strong conservative role in sex determination in eukaryotes (Herpin & Schartl, 2015; Vicoso, 2019).

Alternative splicing has been suggested to play a crucial role in sex determination and gonadal sex differentiation in vertebrates from recent fundings. In sharp contrast to the well-studied sex-specific splicing role of *Dsx* in sex determination in insects, the role of alternative splicing of (master) sex-determining genes in vertebrates has recently been explored, mostly focusing on adult or developmental expression studies and no functional study on isoform level has yet been conducted. For instance, in the Australian beaded dragon (*Pogona vitticeps*), sex-specific splicing of *Nr5a1* (Nuclear Receptor Subfamily 5 Group A Member 1) was suggested to encode for sex determination (Zhang *et al.*, 2022). It hypothesized that the W-linked isoforms can function as a competitive inhibitor to the full-length intact protein to suppress testis development in ZW females (Zhang *et al.*, 2022). For alternative splicing of *Dmrt1*, we have compiled in total 16 vertebrate species (Table 2), 10 fishes, 2 mammals (human and mouse), 4 reptiles including avians (1 chicken, 1 turtle, 1 crocodile and 1 alligator). There are clear interesting patterns. First, it is unclear whether alternative splicing is crucial genetic mechanism determining sex in these vertebrates. For chicken with *Dmrt1* being the master sex-determining gene, it is unclear whether and how splicing variants play a role in sex determination (Zhao *et al.*, 2007). For the remaining species, the master sex-determining gene has not yet been identified or verified. Second, unlike the constitutional sex-specific splicing variants of *Dsx* which generates female and male-specific DSX protein in insects (Hediger *et al.*, 2004), sex-specific splicing variants of *Dmrt1* in vertebrates seem to evolve in a species-specific manner, highly dynamics across tissues and developmental stages, and tend to involve a high number of isoforms (Table 2, Figure 4B). One study showed exclusive expression in testis in both adult and gonadal sex differentiation stages, both showed male-limited expression, suggesting a role in testis formation and function (Raghuveer & Senthilkumaran, 2009). In several (6) species, one or all *Dmrt1* isoforms were exclusively expressed in adult testes or developmental testes (Huang *et al.*, 2005; Lu *et al.*, 2007; Zhao *et al.*, 2007; Deloffre *et al.*, 2009; Su *et al.*, 2015; Domingos *et al.*, 2018). The rest 10 species showed significantly higher expression in adult testis, and during gonadal sex differentiation or transformation from females to males (Table 2). All these expression dynamics suggest a crucial role of *Dmrt1* isoforms in testicular formation and/or function. Lastly, to our best knowledge, so far there have been no reported alternative splicing in any amphibian species. Finally, it is unclear whether all isoforms are functionally involved in sexual differentiation, and whether a higher number of *Dmrt1* isoforms in invertebrates than *Dsx* in insects is due to higher complexity of organism structures. Thus, a broader spectrum of taxa and functional studies are needed to better understand its role in sex determination and sexual differentiation.

### *7.1 Alternative splicing of Dmrt1 in fishes*

Sex-determining mechanism in fishes is extremely diverse and labile, governed by genetic factors, environmental factors, or their interactions. Certain teleost fishes undergo natural sequential sex change, potentially an ideal system to study sex determination and sexual differentiation on which one sex is transformed into the other. Among the 10 fish species with alternative splicing of *Dmrt1*, the isoform number ranges from 2 to 6 (Table 2, Figure 4C). Half of all such species do not show, at any developmental stage, male-specific splicing of *Dmrt1*, and all isoforms showed higher expression in adult testes or developmental testes (Table 2, Figure 4B). For instance, zebrafish (*Danio rerio*) is shown to have ZZ/ZW sex chromosome system with a polygenic sex-determining mechanism with varying contributing loci including *Dmrt1, Amh, Sox9, Cyp19a1 and Foxl2a* (Kossack & Draper, 2019). In this species, alternative splicing has been found in *Dmrt1* with three different isoforms (*Dmrt1a, Dmrt1b, Dmrt1c*) which encoded proteins with 267, 246 and 132 amino acids. All three isoforms are expressed in both testis and ovary but are significantly highly expressed in testis (Guo *et al.*, 2005).

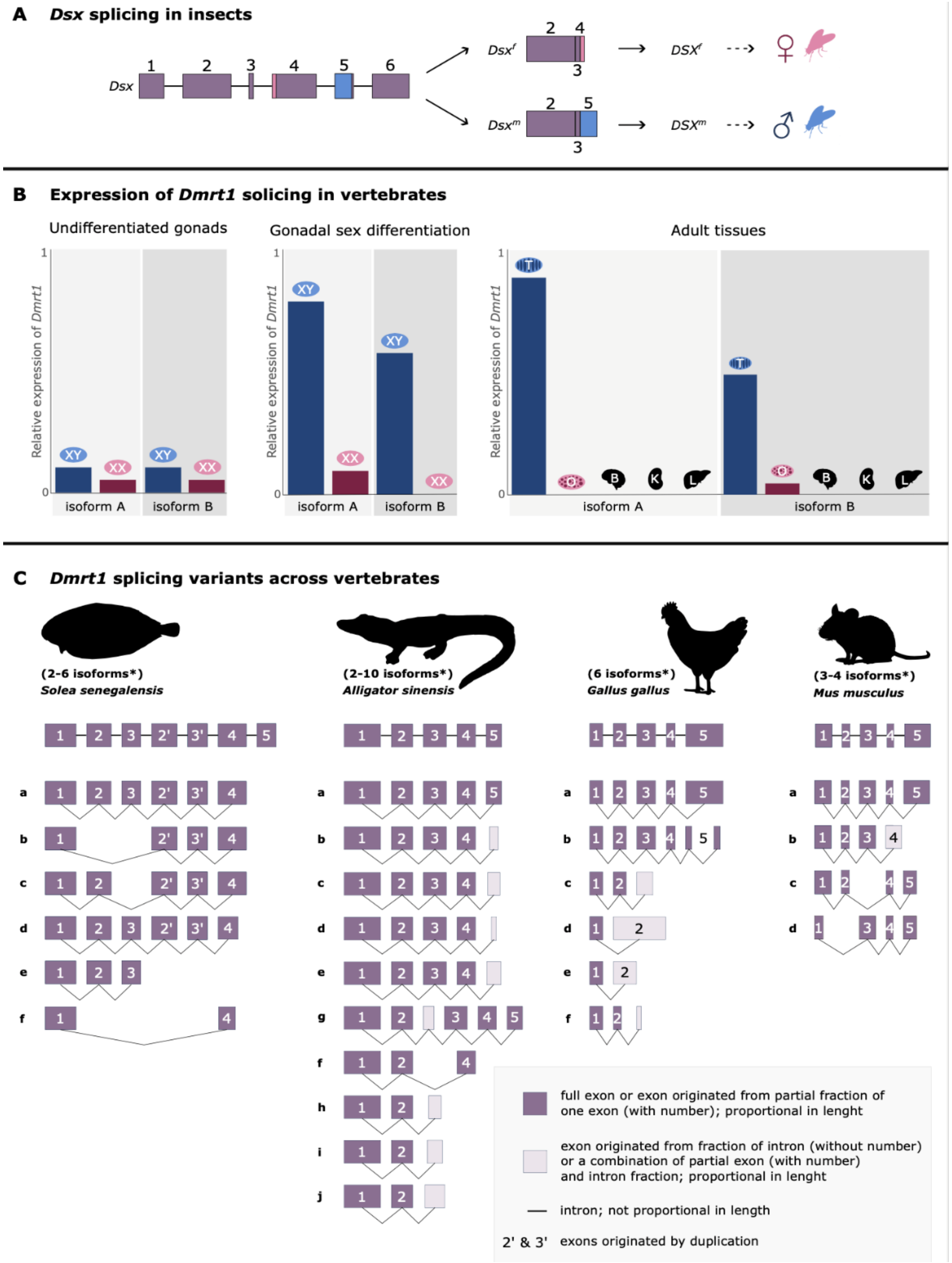

**Figure 4.** Sex-specific splicing of *Dsx* in insects and alternative splicing of *Dmrt1* in vertebrates. A. Sex-specific splicing of *Dsx* in *Drosophila melanogaster* is illustrated and can be found in all insects. Illustration is adapted from (Hediger *et al.*, 2004). The female isoform of *Dsx* mRNAs can produce functional female DSX protein and splicing of male isoform *Dsx* mRNAs generates male DSX protein. B. Summary of expression dynamics of *Dmrt1* alternative splicing isoforms, illustrated by example of 2 splicing isoforms from several vertebrates (Table 2). *Dmrt1* does not have female or male-specific splicing isoforms as insects, instead different isoforms (or all isoforms) could have stage-specific male-limited expression, as illustrated here isoform B has male-limited expression at gonadal sex differentiation stage and isoform A has male-limited expression in adult testis. All isoforms typically have strong male-biased expression during gonadal sex differentiation and in testis than ovary, and often they either do not have or have very weak expression in somatic tissues such as brain (B), kidney (K) and liver (L). In some studies, isoforms also start expressing and with male-biased expression in undifferentiated gonads during embryogenesis. C. One example of illustration of alternative-splicing isoforms from each major vertebrate group except amphibian, and the species from each group was chosen to have the highest number of isoforms found in that group. *: the isoform number range found in that group. Both within and across vertebrate groups, the alternative splicing isoforms do not have the fixed mechanism or pattern, and it seems to be rather random. For example, isoform b in fish is splicing out of exon 2 and exon 3, in reptile is splicing out of exon 5 and retaining a small part of intron region, in chicken is splicing out of exon 5 and retaining both intron 4 and intron 5 region, in mouse is splicing out exon 5 and exon 4 is joining between the entire exon 4 and a fraction of exon 5.

In certain species all isoforms showed testis-specific expression at adult stage but were all expressed in both sexes during gonadal sex differentiation process. For instance, in the rice field eel (*Monopterus albus*) with protogynous hermaphrodite, the 4 isoforms of *Dmrt1* (encoding proteins with 301, 196, 300 and 205 amino acids) were exclusively expressed in adult gonads. Two of the splice forms, *Dmrt1a* and *Dmrt1b*, were expressed during the gonadal transformation from ovary to testis that occurs in the sex change process (Huang *et al.*, 2005). In the European sea bass (*Dicentrarchus labrax*), a teleost fish with polygenic sex determination induced by temperature, two alternative splicing of *Dmrt1* were detected (*Dmrt1a* and *Dmrt1b*), with *Dmrt1a* splicing out exon 4 (Deloffre *et al.*, 2009). Both transcripts were expressed at a similar level and first detected at 150 days post-hatch coinciding with the start of sex differentiation. With development, the expression increased in testis and decreased in ovary, and was only expressed in testis tissues in adults. Both transcripts were suggested to be required for testicular function (Deloffre *et al.*, 2009).

In the fish barramundi (*Lates calcarifer*), a protandrous hermaphrodite species, two isoforms were identified, *Drmt1a* and *Dmrt1b*. Interestingly, different isoforms showed different sex-limited expression (Domingos *et al.*, 2018). The isoform *Dmrt1a*, which composes the DM domain, showed exclusive expression in adult testis, while *Dmrt1b* (lacking this DM domain) was detected in testes and ovaries, with higher expression in testis (Domingos *et al.*, 2018).

*7.2 Alternative splicing of Dmrt1 in mammals*

3 and 4 alternative splicing variants of *Dmrt1* were identified in human and mouse respectively (Figure 4C, Table 2). Their predominant expression in testis indicates a certain role in testis function, yet whether and how each isoform functions is unclear. In human, all 3 *Dmrt1* isoforms were expressed in testis (Cheng *et al.*, 2006). However, sex-specific splicing was not concluded due to expression was only conducted in the testis tissue. In mouse, all 4 *Dmrt1* isoforms showed testis-limited expression at adult stage, but all were expressed during gonadal sex differentiation process with higher expression in males (Lu *et al.*, 2007).

*7.3 Alternative splicing of Dmrt1 in reptiles*

2-10 isoforms of *Dmrt1* have been identified in various reptiles, with the highest number 10 in the alligator *Alligator sinensis* (Figure 4C) (Table 2). In many reptiles, *Dmrt1* expression was upregulated in male embryos. In the painted turtle (*Chrysemys picta*) with TSD, *Dmrt1* was upregulated in males during gonadal development. The study further found that alternative splicings of *Dmrt1* were differentially expressed by temperature at stages 19 and 22 in developing gonads after the onset of sex determination. Both splicing variants displayed a significant male-biased expression pattern when exposed to male-producing temperatures (Mizoguchi & Valenzuela, 2020). In the crocodile *Crocodylus palustris* and the alligator *Alligator sinensis*, 2 and 10 *Dmrt1* isoforms were identified respectively (Figure 4C, Table 2) (Anand *et al.*, 2008; Wan *et al.*, 2013). No isoform was strictly sex specific as they were expressed in both testis and ovary tissues, as well as during gonadal sex differentiation process.

All birds have genetic sex determination with female heterogametic (ZZ/ZW) sex chromosome system. The *Dmrt1* gene has been identified as master sex-determining gene in chicken with dosage effects (Smith *et al.*, 2009). Chicken *Dmrt1* generated 6 different splicing variants during gonadal sex differentiation and their expression was testis-specific in adults. The splicing variants had differential expression along the different stages, suggesting stage-specific splicing variants. *Dmrt1b* was exclusively expressed in embryo gonads and higher in male than female gonads at stage 31 which was the key time of gonadal sexual differentiation. *Dmrt1c* was highly expressed in female than male gonads at stage 31. *Dmrt1f* was detectable only in male gonads at stage 31 (Zhao *et al.*, 2007).

## 8. Conclusion

Nine *Dmrt* genes with conserved DM domain with DNA-binding motif are distributed across vertebrate groups. *Dmrt1* is the most studied gene within *Dmrt* family regarding transcription regulation of DMRT protein and plays a crucial role in regulating sex determination and gonadal differentiation across vertebrates. Our compiled data of *Dmrt1* studies across vertebrate species show that *Dmrt1* is expressed in the early undifferentiated gonads in certain species. Mostly *Dmrt1* shows sexually dimorphic expression, being significantly higher or exclusively expressed, in males during gonadal sex differentiation and predominantly expressed in adult testes, with minimal or no expression in ovaries and other somatic tissues. Notably, *Dmrt1* has been identified as a master or key sex-determining gene in chicken, a frog, a fish and a turtle species, and proposed as a candidate sex-determining gene in additional 12 species (6 fishes, 2 amphibians, 4 birds).

Recent studies highlight the role of epigenetic regulation, involving methylation of the *Dmrt1* promoter region, transposable element insertion and ncRNA involvement, in sex determination and gonadal sex differentiation. Furthermore, alternative splicing of *Dmrt1*, akin to the sex-specific role of *Dsx* in insects, showed lineage-specific isoform number, as high as 10 isoforms in certain species, highly dynamic expressions throughout development and across adult tissues. There seems to be no constitutional sex-specific splicing protein production across all vertebrates, highlighting a major difference in the regulatory mechanism between *Dmrt1* in vertebrates and *Dsx* in insects.

One open question is how generalizable the molecular mechanisms identified in the red-eared slider turtle with ESD system are across vertebrate taxa. Future research should focus on investigating the molecular basis of ESD in a broader spectrum of species. Furthermore, it is crucial to explore the role of epigenetic regulation in the continuum of sex-determining systems, ranging from GSD to systems where GSD interacts with ESD, to pure ESD. Detailed molecular studies are needed to understand these underlying mechanisms. Finally, the genetic mechanism of alternative splicing of *Dmrt1* in sex determination and gonadal sex differentiation has yet been explored and requires further understanding. It is essential to understand why and how multiple (up to 10) alternative splicing variants of *Dmrt1* evolve, whether and if each isoform plays a (specific) role in sex determination and gonadal sex differentiation. Another interesting aspect is the significant differences in the molecular mechanisms and functions of alternative splicing between *Dmrt1* in vertebrates and sex-specific splicing of *Dsx* in insects. Understanding the differences could provide deeper insights into the evolution of sex-determining mechanisms between vertebrates and insects.

**Author Contributions**: W.-J. M. designed the study. B. A. and W.-J. M. collected data and synthesized ideas from the literature. B. A. visualized the synthesized ideas in figures with significant input from W.-J. M. W.-J. M. and B. A. wrote the manuscript. All authors have read and agreed to the submitted version of the manuscript.

**Data Availability Statement**: NA.

**Acknowledgments**: We thank Paris Veltsos for insightful discussion and comments on the manuscript. We also thank two anonymous reviewers for their constructive comments in the earlier version of the manuscript. This work is funded by the ERC starting grant (FrogWY, 101039501) and a starting grant from Vrije Universiteit Brussel (OZR4049) to Wen-Juan Ma. Views and opinions expressed are however those of the author(s) only and do not necessarily reflect those of the European Union or the European Research Council Executive Agency. Neither the European Union nor the granting authority can be held responsible for them.

**Conflicts of Interest**

The authors declare no conflict of interest.

Table 1. List of species across various vertebrate groups which *Dmrt1* was identified as master, or candidate master, sex-determining gene, and the expression pattern throughout gonadal development and at adult stages.

| Class | Species | SD system | Master SD gene | Expression during gonadal sex differentiation | Expression in gonads of sub-adult/adult | Expression in adult somatic tissue | Method | Gene function | Reference |
|---|---|---|---|---|---|---|---|---|---|
| Actinopterygii | *Oryzias latipes* | GSD (XX/XY) | yes (*Dmrt1bY*/*DmY*) | testes (*Dmrt1a, Dmrt1bY/DmY*) | testes (*Dmrt1bY/DmY; Dmrt1a*), ovaries *Dmrt1a* | *Dmrt1bY/DmY* expressed: pituitary, heart, liver, spleen; *Dmrt1bY/DmY* very weak expression (intestine, eye); *Dmrt1a* expressed: spleen (males); *Dmrt1a* very weak expression: spleen (females); not expressed: brain, muscle, gills | PCR, histology | SD, gonad differentiation (*Dmrt1bY*/*DmY*), spermatogenesis (*Dmrt1a*) | Ohmuro-Matsuyama *et al.*, 2003; Kobayashi *et al.*, 2004 |
| Actinopterygii | *Astyanax scabripinnis* | GSD (ZZ/ZW) | candidate | NA | testes | not examined | qPCR | NA | Castro *et al.*, 2019 |
| Actinopterygii | *Cynoglossus semilaevis* | GSD (ZZ/ZW) | candidate | testes | testes | not examined | qPCR | testes differentiation, spermatogenesis | Cui *et al.*, 2017 |
| Actinopterygii | *Betta splendens* | GSD (XX/XY) | candidate | males (testes) | testes | not examined | qPCR | gonad differentiation | Wang *et al.*, 2022b |
| Actinopterygii | *Scatophagus argus* | GSD (XX/XY) | candidate | NA | testes | not examined | PCR | NA | Mustapha *et al.*, 2018 |
| Actinopterygii | *Scatophagus tetracanthus* | GSD (XX/XY) | candidate (*Dmrt1Y*) | NA | tetstes (*Dmrt1X*, *Dmrt1Y*), ovaries (*Dmrt1X* weak) | not examined | qPCR, PCR | NA | Peng *et al.*, 2023 |
| Actinopterygii | *Selenotoca multifasciata* | GSD (XX/XY) | candidate (*Dmrt1Y*) | NA | testes (*Dmrt1X*, *Dmrt1Y*), ovaries (*Dmrt1X)* | not examined | PCR, RNAseq | NA | Jiang *et al.*, 2022 |
| Amphibia | *Hyla arborea* | GSD (XX/XY) | candidate | testes | NA | not examined | histology | NA | Piprek *et al.*, 2013 |
| Amphibia | *Xenopus laevis* | GSD (ZZ/ZW) | yes (*DmW*) | testes (*Dmrt1*), ovaries (*Dmrt1* weak, *DmW*) | testes (*Dmrt1*), ovaries (*DmW* weak, *Dmrt1* weak) | *DmW* expressed: liver; very weak expression mesonephros; *DmW* not expressed: intestine, brain, stomach, spleen, heart | qPCR, PCR, histology, Southern blot | NA | Yoshimoto *et al.*, 2008, 2010; Fujitani *et al.*, 2016; Mawaribuchi *et al.*, 2017b |
| Amphibia | *Rana temporaria* | GSD (XX/XY) | candidate | testes (strong), ovaries (weak) | NA | not examined | histology | NA | Piprek et al. 2013 |
| | *Cairina moschata* | GSD (ZZ/ZW) | candidate | testes (strong), ovaries (weak) | NA | not examined | qPCR, RNAseq | SD, gonad differentiation | Wang *et al.*, 2015b; Bai *et al.*, 2020 |
| | *Dromaius novaehollandiae* | GSD (ZZ/ZW) | candidate | testes (strong), ovaries (weak) | NA | not examined | qPCR | SD, gonad differentiation | Hirst *et al.*, 2017 |
| | *Taeniopygia guttata* | GSD (ZZ/ZW) | candidate | testes (strong), ovaries (weak) | NA | not examined | qPCR | SD, gonad differentiation | Hirst et al. 2017 |
| Reptilia | *Gallus gallus domesticus* | GSD (ZZ/ZW) | yes (*Dmrt1*) | testes (strong), ovaries (weak) | testes | expressed: heart; not expressed: liver, spleen, lung, kidney | qPCR, histology | SD, gonad differentiation, germ cell proliferation | Raymond et al., 1999b; Zhao *et al.*, 2007; Lambeth *et al.*, 2014; Clinton & Zhao, 2023 |
| | *Coturnix japonica* | GSD (ZZ/ZW) | candidate | testes (strong), ovaries (weak) | testes | not examined | RNAseq | SD, gonad differentiation | Okuno *et al.*, 2020 |
| | *Trachemys scripta* | TSD | yes (*Dmrt1*) | testes (strong), ovaries (weak) | NA | not examined | qPCR, histology | gonad differentiation | Bieser & Wibbels, 2014; Ge et al. 2017 |

Table 2. List of vertebrates with alternative splicing of *Dmrt1* and their expression throughout development.

| **Class** | **Species** | **SD system** | **Isoform number** | **Sex-limited expression** | **Expression breath in adult** | **Expression during gonadal sex differentiation or sex change** | **Reference** |
|---|---|---|---|---|---|---|---|
| Actinopterygii | *Clarias gariepinus* | GSD (XX/XY) | 3 | yes | testis | all isoforms are only expressed in developing males | Raghuveer & Senthilkumaran, 2009 |
| | *Danio rerio* | GSD (ZZ/ZW), polygenic | 3 | No | *Dmrt1a* is highly expressed in testis; weaker expression of *Dmrt1b* and *Dmrt1c* | NA | Guo *et al.*, 2005; Kossack & Draper, 2019 |
| | *Megalobrama amblycephala* | GSD (XX/XY) | 4 | yes (adult only) | testis | all isoforms are expressed in males and only one isoform is expressed in initial stage during transformation into female | Su *et al.*, 2015 |
| | *Halichoeres poecilopterus* | Protogynous hermaphrodite | 2 | No | testis and ovary (higher expression in testis) | both isoforms are expressed but higher in males than females | Miyake et al. 2012 |
| | *Lates calcarifer* | Protandrous hermaphrodites | 2 | Yes (*Dmrt1a*), no (*Dmrt1b*) | *Dmrt1a* is only expressed in testis; *Dmrt1b* in both | NA | Domingos *et al.*, 2018 |
| | *Dicentrarchus labrax* | GSD | 2 | yes (only at adult stage) | not in somatic tissues, but in testis | both isoforms are expressed, both are highly expressed in testis | Deloffre *et al.*, 2009 |
| | *Epinephelus merra* | Protogynous hermaphrodite | 2 | No | testis and ovary | all isoforms are expressed in both sexes but higher in transformation into males | Alam *et al.*, 2008 |
| | *Monopterus albus* | Protogynous hermaphrodite | 4 | yes (only at adult stage) | not in somatic tissues, only in testis | all isoforms are expressed in developing gonads in both sexes, significantly higher in testis | Huang *et al.*, 2005 |
| | *Solea senegalensis* | GSD (XX/XY) | 6 | No | testis and ovary (higher in testis) | NA | Cross *et al.*, 2020 |
| | *Acipenser schrenckii* | GSD (ZZ/ZW) | 2 | No | testis and ovary (higher in testis) | both isoforms are expressed but higher in developing females | Okada *et al.*, 2017 |
| Mammalia | *Homo sapiens* | GSD (XX/XY) | 3 | Inconclusive | only studied in testis | NA | Cheng *et al.*, 2006 |
| | *Mus musculus* | GSD (XX/XY) | 4 | yes (only at adult stage) | testis | all isoforms are expressed in developing gonads in both sexes, significantly higher in males | Lu *et al.*, 2007 |
| Reptilia | *Crocodylus palustris* | TSD | 8 | No | testis, kidney, brain | all isoforms are expressed throughout temperature sensitive period, higher in males than females | Anand et al., 2008 |
| | *Chrysemys picta* | TSD | 2 | No | testis and ovary (higher in testis) | both isoforms are expressed but higher in males than females | Mizoguchi & Valenzuela, 2020 |
| | *Alligator sinensis* | TSD | 10 | No | testis and ovary | NA | Wan *et al.*, 2013 |
| | *Gallus gallus* | GSD (ZZ/ZW) | 6 | yes (only testis at developing stage 31) | testis and heart | *Dmrt1a* is not expressed; *Dmrt1b, Dmrt1c, Dmrt1d, Dmrt1e* are expressed in both sexes; *Dmrt1f* is only expressed in testis | Zhao et al. 2007; |